\documentclass[twocolumn,showpacs]{revtex4-1}

\usepackage{graphicx}
\usepackage{float, amsmath, amssymb, color}

\begin{document}

\title{Breathing mode in the Bose-Hubbard chain with a harmonic trapping potential}

\author{Wladimir Tschischik}
\author{Roderich Moessner}
\author{Masudul Haque}

\affiliation{Max-Planck-Institut f\"ur Physik komplexer Systeme, N\"othnitzer Str. 38, 01187
  Dresden, Germany} 

\date{\today}

\begin{abstract}

We investigate the breathing mode of harmonically trapped bosons in an optical lattice at small site
occupancies.  The Bose-Hubbard model with a trapping potential is used to describe the
breathing-mode dynamics initiated through weak quenches of the trap strength.  We connect to results
for continuum bosons (Lieb-Liniger and Gross-Pitaevskii results) and also present deviations from
continuum physics.
%
%
We take a spectral perspective, identifying the breathing mode frequency with a particular energy
gap in the spectrum of the trapped Bose-Hubbard Hamiltonian.  We present the low energy
eigenspectrum of the trapped many-boson system, and study overlaps of the initial state with
eigenstates of the quenched Hamiltonian.  There is an intermediate interaction regime, between a
"free-boson" limit and a "free-fermion" limit, in which the Bose-Hubbard breathing mode frequency
approaches the Gross-Pitaevskii prediction. In addition, we present a striking failure of the 
time-dependent Gutzwiller approximation for describing breathing modes.

\end{abstract} 

\maketitle 

\section{Introduction}

Controlled and tunable experimental realizations of confined quantum systems with ultra-cold atoms
enables non-equilibrium studies of quantum many-body states in novel geometries.  A common feature
of many cold-atom experiments is the confinement of a many-body system in a harmonic trap.  Trapping
introduces many distinctive features which have no analog in uniform many-body states, such as
collective excitations like breathing modes, dipole modes, and scissors modes.  Such trap-related
collective modes have been widely studied both experimentally and theoretically for continuum
systems, especially in the mean-field regime, since the early days after quantum degeneracy was
achieved with trapped atoms \cite{early_BEC}.
In a well-known recent experiment \cite{Naegel_exp}, breathing modes have been used for a
one-dimensional (1D) system of continuum bosons to characterize mean-field and non-mean-field regimes, as
well as states obtained by quenching to large negative interactions.
With the addition of optical lattices, it should be possible to study collective modes in
lattice systems beyond the mean field regime.
In this work, we study the breathing mode for interacting bosons on a 1D lattice, described by the
Bose-Hubbard model and subject to an additional trapping potential.

Our work addresses the dynamics of a finite number ($N$) of bosons on a 1D chain with $L>N$ sites,
subject to the Hamiltonian
\begin{multline}
 H_{BH} ~=~ - J \sum_{i=1}^{L-1} \left( b_{i}^\dagger b_{i+1}  +  b_{i+1}^\dagger b_{i}  \right)
\\  ~+~  \frac{U}{2} \sum_{i=1}^{L}  \hat{n}_{i}(\hat{n}_{i}-1)   ~+~ \sum_{i=1}^{L} V(i)\hat{n}_{i}  
\,  .
\label{eq:Hamiltonian}
\end{multline}
Here $b_{i}$, $b_{i,}^\dagger$ are the bosonic operators for the site $i$ ($ i = 1 \ldots L$), and
$\hat{n}_{i}=b_{i}^\dagger b_{i}$.  We will measure energy [time] in units of the tunnel coupling
$J$ [inverse tunnel coupling $1/J$], and therefore set $J=\hbar=1$.  The trapping potential
\begin{equation}
V(i) ~=~ \frac{1}{2}k\left( i- \frac{L+1}{2}\right)^2  
\end{equation}
is centered at the midpoint of the chain.

\begin{figure}[tb]
\centering
\includegraphics[width=.9\columnwidth]{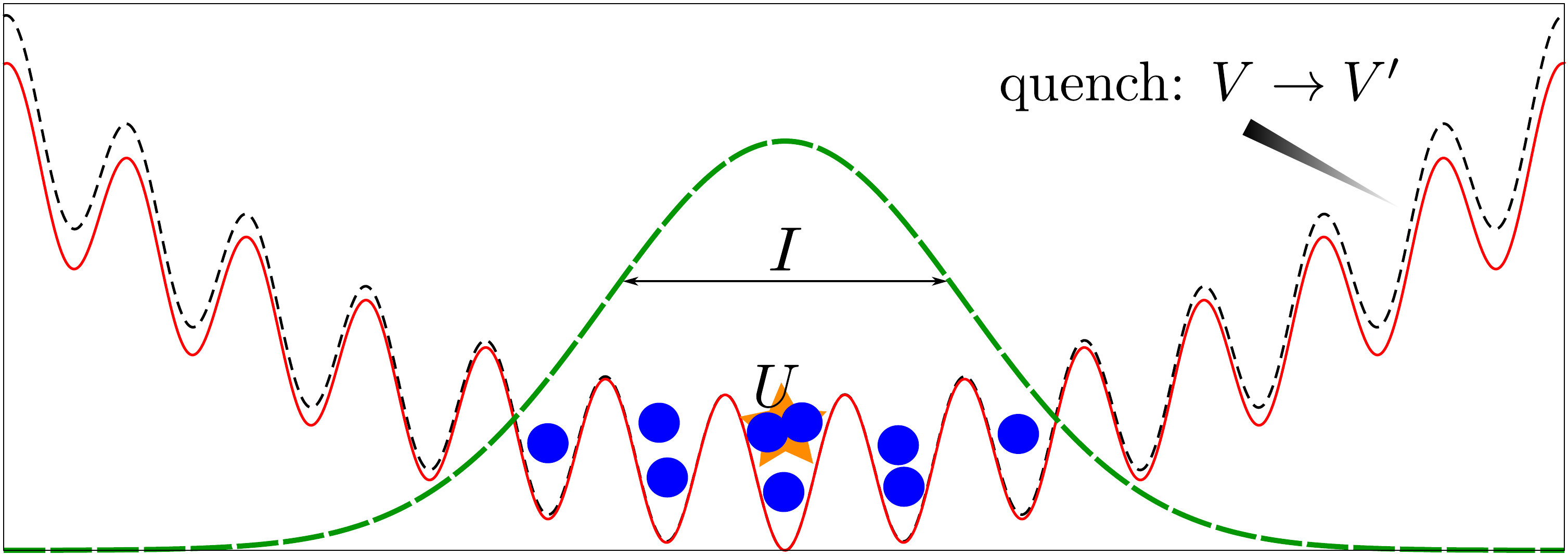}
\caption{(color online) Schematic of bosons in an optical lattice and overall harmonic trapping
  potential.  The trap strength is quenched from $V$ to $V'$.  The bell-shaped (dashed green) curve
  represents the density profile, shown schematically as a continuous curve; in our tight-binding
  description it is a discrete function (site occupancies).
} 
\label{fig:breathingModeExplanation}
\end{figure}

The breathing mode can be excited by a quench (sudden change) of the trapping potential strength $k$.
Fig. \ref{fig:breathingModeExplanation} shows schematically lattice bosons trapped in a harmonic
trap and a quench of the trapping strength $k$, i.e. a reduction of  harmonic confinement.  In
subsequent time evolution the bosonic cloud undergoes expansion and contraction dynamics.  We
characterize this dynamics using the second moment of the density distribution $I = \sum_i \hat{n}_i
(i-\frac{L+1}{2})^2$, which measures the cloud width.  After a quench the second
moment shows oscillatory behavior.  If the dynamics $I(t)$ is nearly monochromatic, the dominant
frequency $\Omega$ can be identified as the breathing mode frequency.  The dynamics is expected
to be nearly monochromatic if the quench is small; we will concentrate on small quenches of $k$.

Collective modes in 1D Bose-Hubbard systems have been previously addressed in Refs.\ \cite{Lundh,
  Peotta_DiVentra_2013, exactSolutionTrap, Snoek_PRA12}.
We focus on the connection to and deviations from continuum physics, and hence concentrate on
situations where the site occupancy is everywhere lower than unity.  This precludes Mott physics
from the present study.  
When the density is low enough, we can approximate the cosine dispersion of a lattice particle by a
quadratic dispersion; this ``effective mass approximation'' $m^*=\frac{1}{2J}$ ascribes a continuum
mass to lattice particles, so that we can relate our trapping strength $k$ to the trapping frequency
of a continuum trapping potential $\frac{1}{2}m\omega_0^2x^2$:
\begin{equation}
 \omega_0 = \sqrt{2kJ}.
\end{equation}
We will take this to be the \emph{definition} of the trapping frequency $\omega_0$ on the lattice,
even when the densities are not small enough for the effective mass approximation to hold.

We will compare with results and approximations from the continuum case.  The continuum limit of the
Bose-Hubbard model is the Lieb-Liniger model
\begin{multline}
 H_{LL} ~=~ -\frac{\hbar^2}{2m}\sum_{j=1}^N \frac{\partial^2}{\partial x_j^2} 
~+~  c\sum_{\langle i,j\rangle} \delta(x_i - x_j) \\ ~+~ \frac{1}{2} m\omega_0^2\sum_i x_i^2 \; , 
\end{multline}
where we have added a trap term.  Full quantum calculations of Lieb-Liniger dynamics in harmonic
traps may be possible using the methods of Refs.~\cite{Caux_Konnik_several}, where related
situations have been studied.  Refs.~\cite{ZollnerSchmelcher_PRA07, Brouzos,
  KroenkeSchmelcher_arxiv2013, CederbaumAlon_arxiv2013} have recently used Multi-Configuration
Time-Dependent Hartree approaches to calculate breathing dynamics and/or eigenspectra of
harmonically trapped continuum bosons in a trap; the lattice results of the present work complement
these continuum results.  The breathing mode of harmonic-trapped 1D continuum bosons has also been
addressed through hydrodynamic ideas, kinetic equations, lattice approximations, and sum rules
\cite{Menotti,Astrakharchik,Kimura_PRA02, FuchsLeyronasCombescot_2003_2004,
  Kraemer_Pitaevskii_Stringari_PRL02, Peotta_DiVentra_2013, Mazets_EPJD11}.  The breathing mode for
trapped 1D continuum bosons has been experimentally measured \cite{Esslinger_exp,Naegel_exp}.

When bosons form a Bose-Einstein condensate, the dynamics of the condensate is well-described in the
continuum through the time-dependent Gross-Pitaevskii (GP) equation
\begin{equation}
i\partial_t \Psi(r,t) =
\left(-\frac{\hbar^2}{2m} \Delta + V(r) + g|\Psi(r,t)|^2\right)\Psi(r,t).   
\end{equation}
The GP or `hydrodynamic' description predicts breathing mode frequencies $\Omega_{GP} = \omega_0
\sqrt{D+2}$, for $D$ dimensions \cite{Stringari}, i.e., $\Omega_{GP} = \sqrt{3}\omega_0$ for 1D, with
weak dependence on the interaction parameter $g$.

The Gross-Pitaevskii description is reliable for higher dimensions, small temperatures, and weak
interactions.  Since there is no true condensation for 1D bosons, one might wonder whether the
prediction has any relevance to the 1D system we are studying.  The experiment of
Ref.~\cite{Naegel_exp} has found a regime of interactions where the breathing-mode frequency
approaches the Gross-Pitaevskii prediction.  The lattice calculations presented in this article,
and the continuum calculations of Ref.\ \cite{KroenkeSchmelcher_arxiv2013}, show
that, even for  true 1D bosons having no transverse degrees of freedom whatsoever, there is a range 
of interactions where the breathing frequency approaches  $\sqrt{3}\omega_0$. 

We will pay particular attention to the spectrum.  Since the many-body system in a trap is a paradigmatic
situation in cold-atom experiments, a thorough description of many-body eigenspectra in the presence
of traps is clearly of fundamental interest \cite{HaugsetHaugerud_PRA98, Brouzos,
  KroenkeSchmelcher_arxiv2013, ZollnerSchmelcher_PRA07, CederbaumAlon_arxiv2013}.  We provide a description of the eigenspectra of 1D many-boson
systems in harmonic traps and relate spectral properties to the breathing mode excitation.  This
description applies qualitatively both to the 1D Bose-Hubbard model as a function of the on-site
interaction, and to the Lieb-Liniger model as a function of the contact interaction.

The breathing mode frequency can be identified as the excitation energy of the lowest spatially
symmetric many-body eigenmode that is excited in a trap quench.  For small enough quenches, we can
neglect the occupation of all but one of the excited states.  If the ground state and this excited
state have energies $E_0$ and $E_n$, the wavefunction evolves as $|\psi(t) \rangle = a_0 e^{-iE_0t}
|\phi_0\rangle + a_n e^{-iE_nt} |\phi_n\rangle$.  As a result, any observable, including the cloud
size, will have oscillation frequency $\Omega=E_n-E_0$.  In Sections \ref{sec:spectrum} and
\ref{sec:breathing_mode} we will identify the energy level relevant for a small trap quench; we
provide a brief summary here.

For non-interacting trapped bosons ($U=0$) in the continuum, the first excited state is at energy
$\hbar\omega_0$ and spatially asymmetric.  There are two degenerate, spatially symmetric, states at
energy $2\hbar\omega_0$.  The same situation occurs in the $U\to\infty$ limit where the bosons act
as free fermions (Tonks-Girardeau limit).  One of these degenerate levels stays flat at
$2\hbar\omega_0$ for all $U$.  In a spatially symmetric trap quench, however, it is the other level
of this pair that gets dominantly excited.  The breathing mode frequency as a function of $U$ can be
identified with the excitation energy of this level, which goes to $2\omega_0$ in the limits
$U=0,\infty$.  At finite $U$, we show that this frequency (energy level) drops below $2\omega_0$,
and the minimum value approaches $\sqrt{3}\omega_0$ as the particle number is increased.  Of course,
this description is strictly valid only in the continuum; for the Bose-Hubbard model there are
deviations at finite densities.

In addition to the exact calculations for spectra and breathing frequencies, we have employed the
time-dependent Gutzwiller approximation for the Bose-Hubbard model.  Despite the popularity of this
time-dependent mean-field approximation, regimes where this might give incorrect dynamics are not
widely known.  We have found that this approximation gives \emph{qualitatively} incorrect results
for the breathing frequency at large interactions.

The spectrum and the breathing mode frequency, as functions of particle number and density, are
described in Sections \ref{sec:spectrum} and \ref{sec:breathing_mode}.  Section \ref{sec:Gutzwiller}
describes the Gutzwiller approximation and its failure to describe breathing mode frequencies at
large $U$.  In Section \ref{sec:strongtrap} we present some features of strong traps, farther from
the regime where the continuum approximation is applicable.

\begin{figure}[tb]
\centering
\includegraphics[width=1\columnwidth]{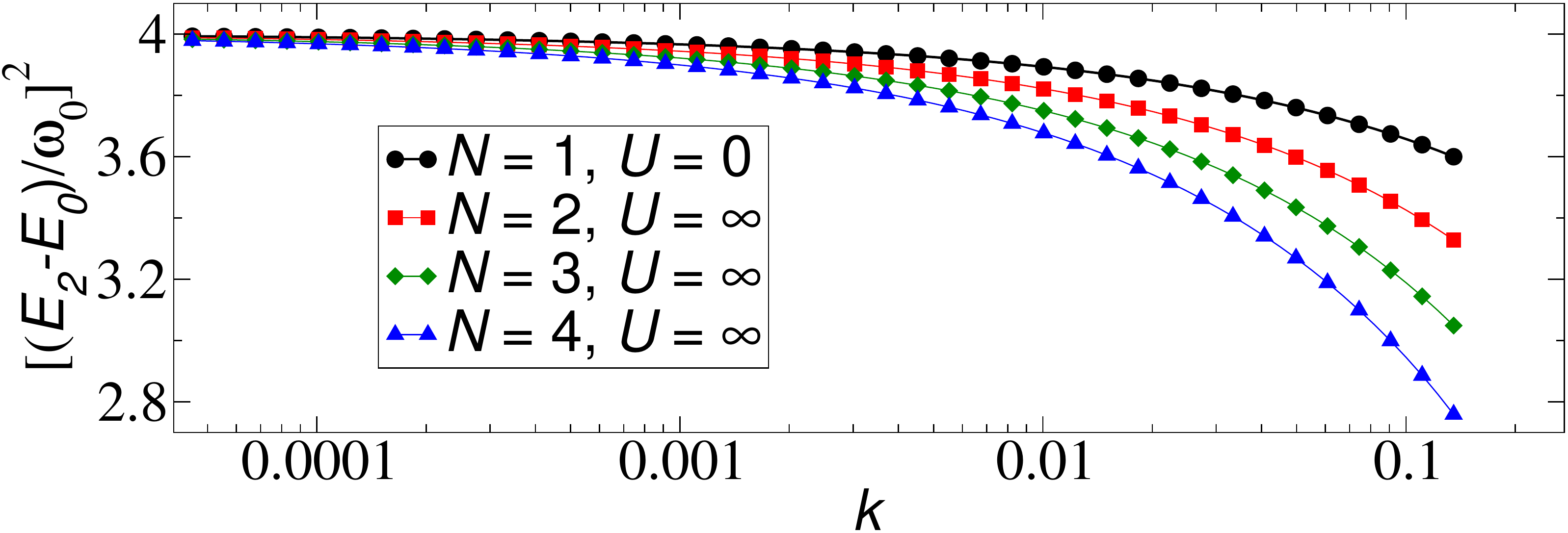}
\caption{ The second excitation energy (breathing mode frequency, $E_2-E_0$) at the free-boson
  ($U=0$) and free-fermion ($U=\infty$) points.  The excitation energy is equal to  $2\omega_0$ in
  the continuum limit  (small $k$) and deviates significantly for stronger traps.
} 
\label{fig:EigenstatesOscillatorsingle_particle}
\end{figure}  

\section{Spectrum of a Bose-Hubbard chain with a harmonic trap \label{sec:spectrum}}   

We describe below the many-boson spectrum for a Bose-Hubbard chain at low filling in the presence of
a harmonic trap, both for the integrable points ($U=0$ and $U=\infty$, subsection
\ref{sec_spectrum_free_points}) and for general $U$ (subsection \ref{sec_spectrum_finite_U}).  We
first describe continuum expectations and then show how the finite-density lattice system deviates.
The continuum case (1D trapped bosons) has been discussed in the literature previously
\cite{HaugsetHaugerud_PRA98, Brouzos, KroenkeSchmelcher_arxiv2013, CederbaumAlon_arxiv2013}.

\subsection{$U=0$ (free bosons) and $U=\infty$ (``free fermions'') \label{sec_spectrum_free_points}}

In the limits $U=0$ (non-interacting bosons) and $U=\infty$ (hard-core bosons, Tonks-Girardeau
limit), the many-body spectrum can be constructed out of the single-particle spectrum.  

In the continuum, the single-particle spectrum is the equally spaced harmonic oscillator spectrum.
The free bosonic ground state is then constructed by placing all bosons in the single-particle
ground state, and excited states are constructed by promoting bosons to higher single-particle
states.  For large $N$, the number of ways one can distribute the $N$ particles in single-particle
levels to get a many-body excited state of excitation energy $E_i-E_0 = l\omega_0$ is the number of
integer partitions $p(l)$ of the integer $l$ \cite{integer_partitions}.  For finite $N$, the
degeneracy sequence of the many-body spectrum is modified to the sequence $p_N(l)$ given by the
number of partitions of $l$ into a maximum of $N$ parts \cite{restricted_integer_partitions}.

The free fermion (hard-core boson) ground state is constructed by filling up the lowest $N$
single-particle states.  Therefore, in the continuum the free fermion ground state energy is larger
than the free-boson ground state energy by $\frac{1}{2}N(N-1)\omega_0$.  The degeneracy counting of
excitations over this ground state is the same as in the free-boson case.

In the presence of a lattice, the single-particle spectrum is modified \cite{exactSolutionTrap} and
no longer equally spaced.  Figure \ref{fig:EigenstatesOscillatorsingle_particle} shows the
excitation energy of the second excited level for lattice bosons.  For $U=0$, the excitation energy
is the same for any number of bosons, and is equal to the excitation energy of the second
single-particle state.  (The many-body excited state is formed by promoting a single particle from
the condensate to this excited level.)  For $U=\infty$, the many-body excited eigenstate is
constructed by promoting a particle across the ``Fermi surface''; hence the second many-body
excitation energy is the difference between the $N$-th and $(N+2)$-th single-particle level, and
thus depends on the number of particles.  Figure \ref{fig:EigenstatesOscillatorsingle_particle}
shows that the deviation of $E_2-E_0$ from the continuum value $2\omega_0$ is smallest for weak
traps, when the central occupancy is low, and increases with increasing $k$.  For large $k$, the
bosonic cloud is localized on few lattice cites.  Such situations are not closely related to
continuum physics and are described in Section \ref{sec:strongtrap}.

For low densities, the $U=0$ ground state is expected to be ${\sim}N\frac{1}{2}\omega_0$ above the
bottom of the band, which is at energy $-2JN$.  This is seen to be approximately true for $N=3$
bosons and $\omega_0=\sqrt{2k}\approx0.0447$ in Figure \ref{fig:spectrum1}a.  In the same figure we
also see the difference between the bosonic and fermionic ground states to be approximately
$\frac{1}{2}N(N-1)\omega_0$.  

For finite chains, the additional confinement of the chain edges also affects the spectrum.  The
effect is stronger for higher excited states and large $U$ because the relevant eigenstates are spatially
more extended.  As seen in Figure \ref{fig:spectrum1}b,
the effect is to increase the excitation energies.

\subsection{Finite $U$ \label{sec_spectrum_finite_U}}

\begin{figure}[t]
\centering
  	\includegraphics[width=1\columnwidth]{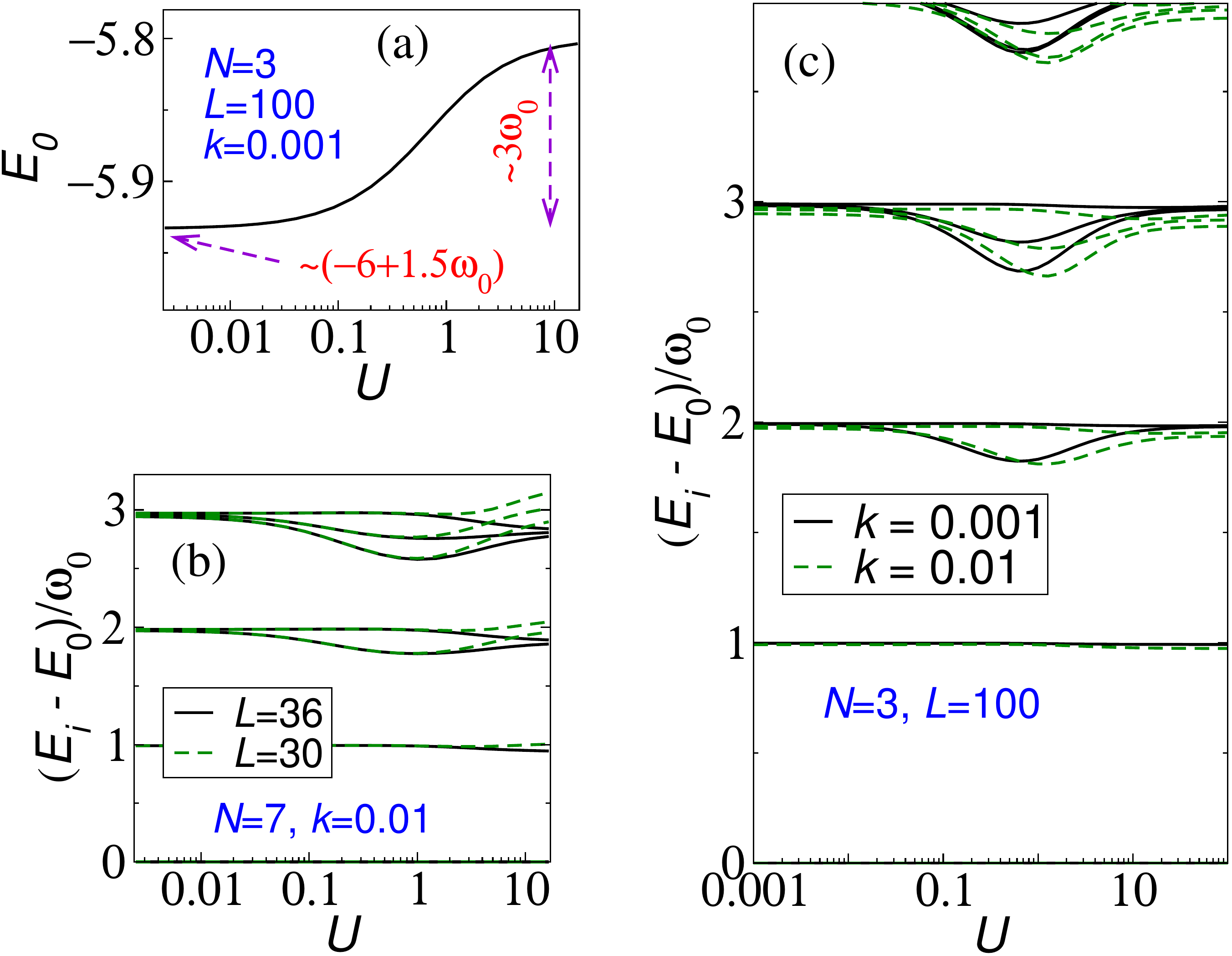}
\caption{  \label{fig:spectrum1}  
Low-energy spectrum of the 1D Bose-Hubbard model with a trap.  
(a) Ground state energies for $N=3$ bosons with weak trapping. 
(b) $N=7$ bosons; comparison of two chain lengths shows the effect of finite edges on the excitation
energies.
(c) $N=3$ bosons; comparison of two trapping potentials shows the deviation from continuum with
stronger traps. 
}
\end{figure}

We first describe the spectrum in the continuum.  

The ground state energy increases monotonically with $U$, changing by $\frac{1}{2}N(N-1)\omega_0$ as
$U$ is varied from $U=0$ to $U=\infty$.  Since we are interested in frequencies observed in
real-time dynamics, we will focus on the excitation energies.  

A prominent feature of the excitation spectrum is that, at each $l\omega_0$, there is a single
energy level that remains constant as a function of $U$.  These are related to the dipole
oscillation (Kohn) mode \cite{KohnMode}.  Introducing center-of-mass and relative coordinates it is
possible to show that the inter-particle interaction appears only in the Hamiltonian describing the
relative motion sector.  Since the center of mass dynamics of the system is independent of the
interaction, an equally spaced harmonic oscillator spectrum appears within the full many-body
spectrum at any $U$.  Formally, there exists a collective ladder operator, commuting with the
interaction term \cite{KohnMode}, which creates a tower of equally spaced eigenenergies separated by
$\omega_0$, independently of the interaction.

At each $l\omega_0$ with $l>1$, the eigenenergies other than the flat level vary
non-monotonically with $U$, dipping below $l\omega_0$ at some finite $U$.  We will associate the
lowest of these non-constant excitation energies with the breathing mode frequency.

Figure \ref{fig:spectrum1}b,c displays this structure for the lowest few levels.  Since our calculations
are on a lattice, there are (small) deviations from the continuum case, which are characterized in
Figure \ref{fig:spectrum1}c by comparing two different trap strengths.  (Smaller $k$ corresponds more
closely to the continuum.)  For a lattice system, the exact degeneracies at $U=0$ and $U=\infty$ are
spoiled, and the levels related to the dipole mode are not completely independent of $U$.  Lattice
effects push the energies below their continuum values.

\section{Breathing mode frequency \label{sec:breathing_mode}}

\begin{figure}[t]
\centering
  	\includegraphics[width=1\columnwidth]{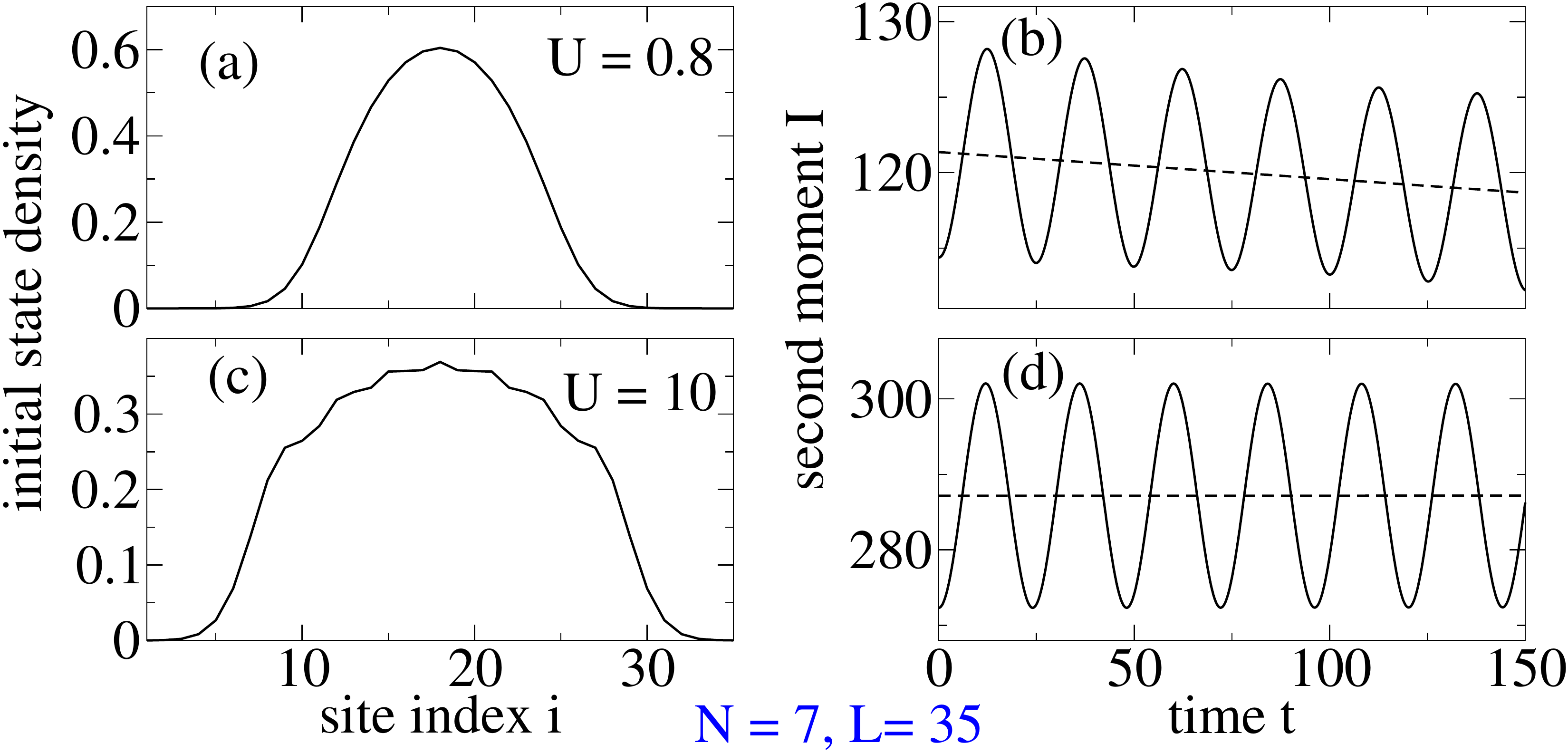}
\caption{ \label{fig:breathin_oscillation_N7} 
(a,c) Ground state density profiles for $N=7$ bosons
  on $L=35$ lattice sites at $k = 0.011$ and (a) $U=0.8$ and (c) $U=10$.  For small $U$ density
  distribution is Gaussian-like; for large $U$ the density
  distribution shows oscillations (fermionization).
(b,d) Breathing mode oscillation after a quench of trapping strength $k = 0.011 \rightarrow 0.01$.
  (b) For $U=0.8$ the influence of a second frequency is seen through the overall decrease in
  addition to oscillations.  (d) For $U=10$ there is no noticeable influence of additional
  frequencies.
}
\end{figure}

In this section we show the connection of the breathing mode frequency $\Omega$ to the low energy
spectrum, by analyzing the overlaps of the initial state with the final eigenstates in quenches of
the trap strength (\ref{sec:overlaps}).  We then describe the dependence of the breathing mode frequency on the
interaction strength $U$ (\ref{sec:non_monotonic_U_dependence}).

\subsection{Quench dynamics \label{sec:quenchdynamics}}

We excite breathing modes through trap quenches.  The initial state is the ground state of an initial
Hamiltonian $H(k_i)$ with trap strength $k_i$.  At time $t=0$ the trapping strength is suddenly
reduced to $k < k_i$; the trapped bosonic cloud then undergoes breathing mode oscillations.  We
perform small quenches, reducing the trapping strength by $5\%$ to $10\%$.

Fig. \ref{fig:breathin_oscillation_N7}(a,c) show the initial density distributions for $N=7$ bosons with
$k=0.011$ and $U=0.8$ ($U=10$).  The density profiles are near-Gaussian for small interactions.  At
large $U$ fermionization is discernible as peaks (oscillations) of the density distribution
\cite{BongsSengstockPfannkuche_PRA07}.  

In Figs.\ \ref{fig:breathin_oscillation_N7}(b,d) we show the time evolution of the second moment of
the cloud after the trap strength is quenched from $k=0.011$ to $k=0.01$.  There is a clear dominant
frequency of oscillation.  We find it natural to identify this dominant frequency as the breathing
mode frequency.  At large $U$, no other feature is visible at these timescales.  At small
$U$, however,  the envelope of the oscillating $I(t)$ has an overall decrease, indicating that at
least two different frequencies are present in the time evolution.

\subsection{Overlaps with eigenstates of the quenched Hamiltonian \label{sec:overlaps}}

\begin{figure}[t]
\centering
  	\includegraphics[width=1\columnwidth]{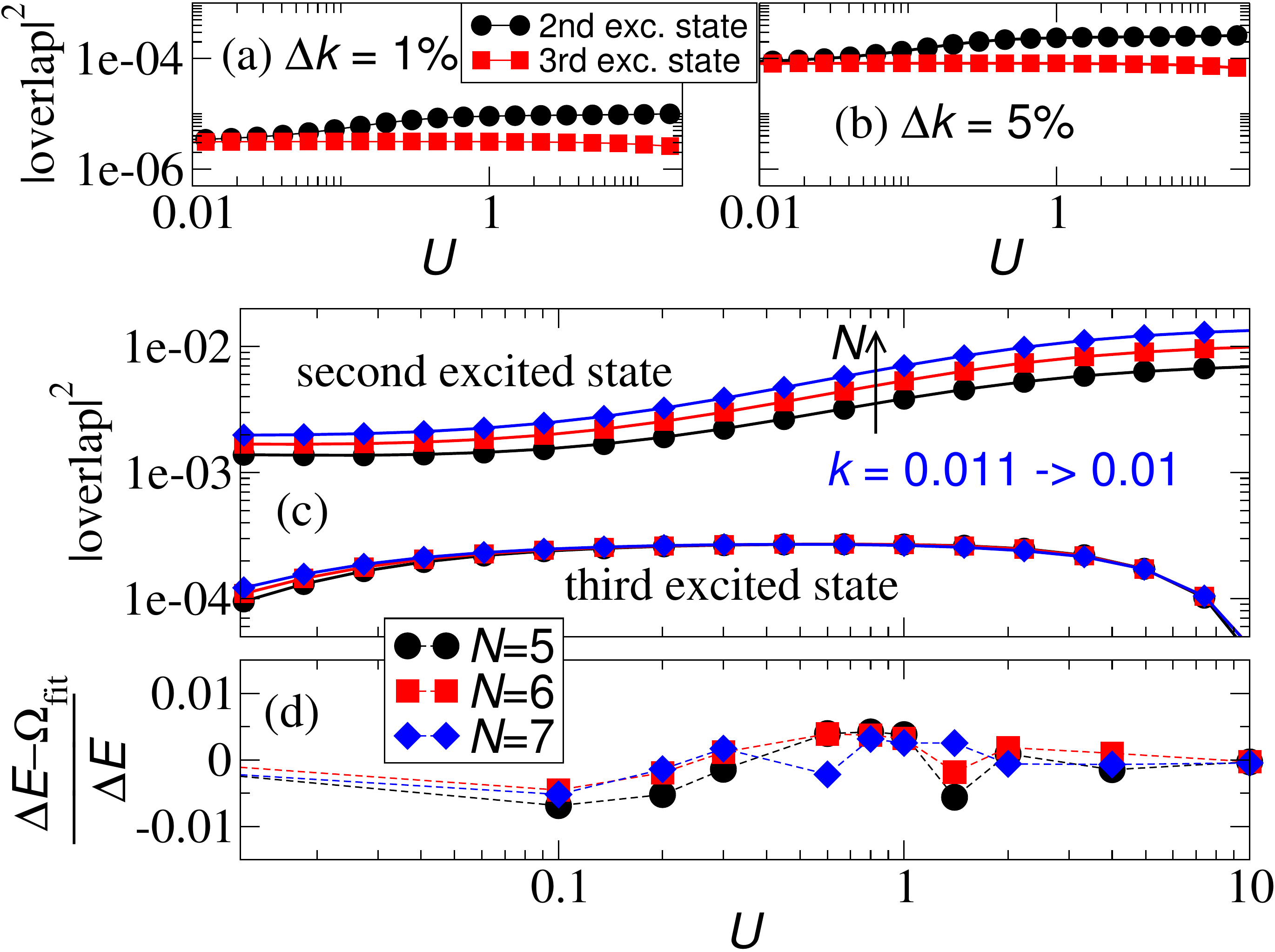}
\caption{(color online) 
(a-c) Overlaps of the initial state with second and third exited states after a quench of trapping
  strength $k$.
(a,b)  $N=2$ particles.  1\% and 5\% reductions of $k$ starting from $k=9\times10^{-6}$.  The
  separation between 2nd and 3rd eigenstates at small $U$ is not reliable because of
  near-degeneracy.     
(c) Overlaps for $N=5,6,7$. Difference between two overlaps increases with $N$.  (d) Relative
  difference between $\Delta E = E_2-E_0$ and breathing frequency $\Omega_{fit}$ obtained by a
  sinusoidal fitting of the oscillating second moment $I(t)$. }
\label{fig:overlaps_N567}
\end{figure}

For small quenches the initial state is a superposition of only the lowest eigenstates of the quenched
Hamiltonian.  The largest overlap (near unity) is with the new ground state.  The first eigenstate is not excited
because of reflection symmetry (it is odd under spatial parity, as opposed to the initial state).
We therefore concentrate on the second and third excited states, as the eigenstates higher than that
have very small weight.

In Figure \ref{fig:overlaps_N567}(a,b) we show the case of two bosons.  For $k<10^{-5}$, the
occupancies are small and the effective mass approximation works well, so that these can be
considered to be continuum results.  The overlap with the second excited state is seen to be
consistently larger than the overlap with the third excited state, except at very small $U$.  At
small $U$, these two excited states are nearly degenerate (c.f., Fig.\ref{fig:spectrum1}), so the
relative values of overlaps in this interactions regime does not affect the breathing frequency.
Comparison of panels (a) and (b) shows that the relative magnitudes of these two overlaps (second
and third excited states) overlaps do not depend strongly of the quench strength, but the absolute
magnitudes increase for larger quench strengths, at the expense of the overlap with the new ground
state.

The difference between overlaps with the second and third excited states increases with increasing
particle number, as shown in Figure \ref{fig:overlaps_N567}c for $N=5,6,7$, for the fixed trap
quench $k=0.011 \to 0.01$.  With $N=7$ particles, the overlap for the second excited state is larger
by more than an order of magnitude than the overlap with the third excited state, over the entire
range of interactions. 

The overlap results indicate that we can identify the excitation energy of the second excited state
as the breathing mode frequency, and that this identification should get better at larger sizes.
This is verified by comparing the the excitation energy $E_2-E_0$ with the frequency of oscillation
of the second moment $I(t)$, as shown in Figure \ref{fig:overlaps_N567}(d).

The relative magnitudes of the overlaps also explain the $I(t)$ behaviors in Figures
\ref{fig:breathin_oscillation_N7}(b,d).  At smaller $U$ the third excited state is not completely
negligible, and therefore the dynamics, although dominated by the breathing frequency
$\Omega=E_2-E_0$, also contains the frequencies $E_3-E_0$ and $E_3-E_2$.  The resulting ``beating''
explains the overall decrease of the oscillating $I(t)$, visible at small $U$ [Figure
  \ref{fig:breathin_oscillation_N7}(b)].  At larger $U$ the relative contribution of the third
excited state is smaller [Figure \ref{fig:overlaps_N567}(a,b,c)], which explains why almost no
beating is visible in Figure \ref{fig:breathin_oscillation_N7}(d).

\subsection{Non-monotonic dependence of breathing frequency on $U$ \label{sec:non_monotonic_U_dependence}}

\begin{figure}[t]
\centering
  	\includegraphics[width=1\columnwidth]{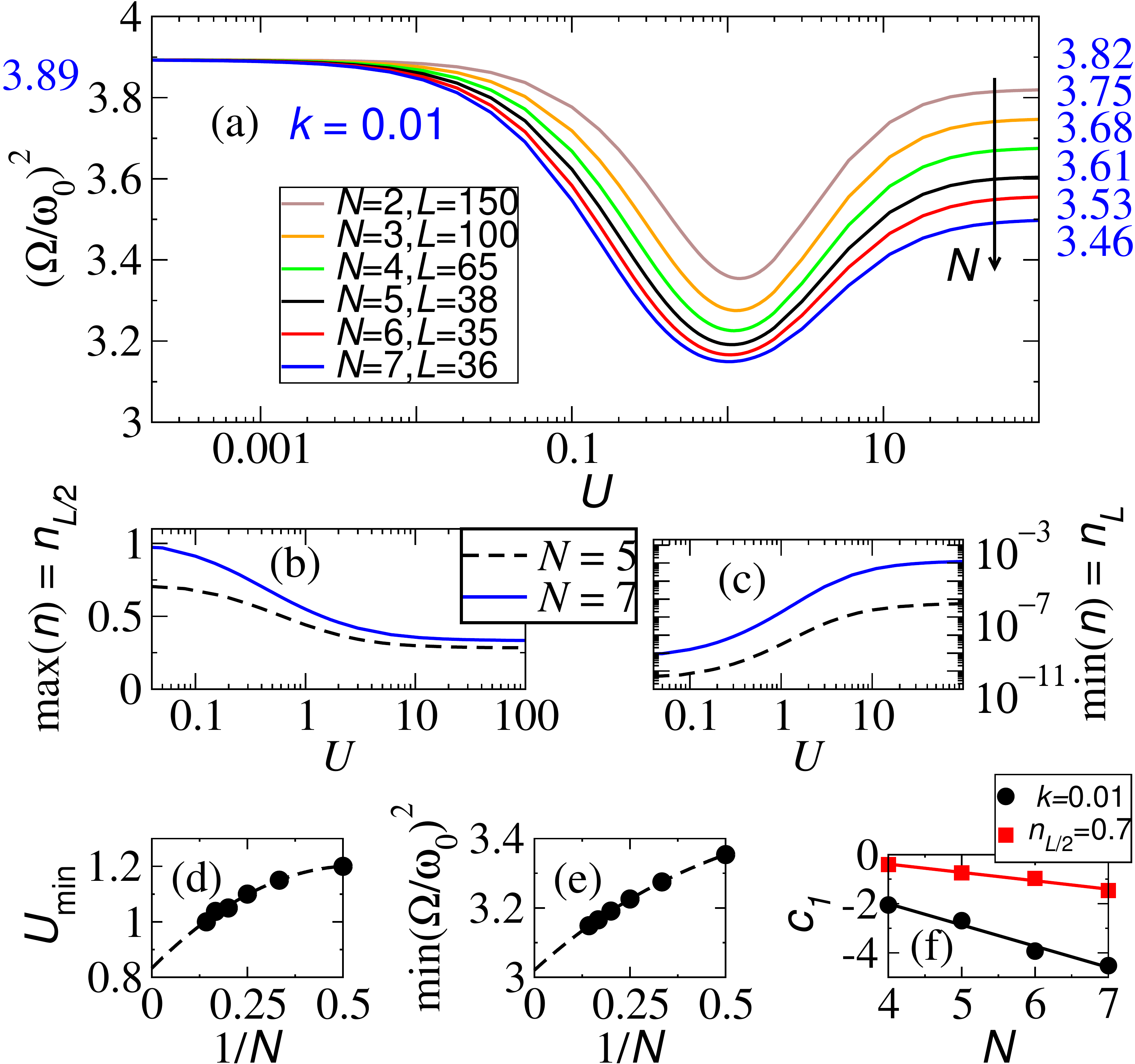}
\caption{(color online) (a) Breathing frequency $\Omega$ as a function of interaction $U$, at a
  fixed trapping strength $k=0.01$.  The numbers on the left (right) show $U=0$ ($U=\infty$) values
  at large $L$.  (b,c) Maximum and minimum of  ground state density, for $N=5$, $L=38$ (full curve)
  and $N=7$, $L=35$ (dashed curve), at
  $k=0.01$.  
(d) Value of the interaction $U$ at which $\Omega$ has a minimum.  
(e) Minimum value of $(\Omega/\omega_0)^2$. In (d,e), Dashed line is a third-order polynomial
  extrapolation to $N\rightarrow \infty$.  (f) Slope of $\Omega(U)$ at small $U$, for a constant
  trapping strength $k=0.01$ and constant central occupancy $n_{L/2} \approx 0.7$.   }
\label{fig:nonmonotonicOmega}
\end{figure}

Having identified the breathing mode frequency $\Omega$ as the energy difference between the ground
and second excited states, $\Omega=E_2-E_0$, in this subsection we describe the dependence of
$\Omega$ on the on-site interaction $U$.  As in the continuum case
\cite{KroenkeSchmelcher_arxiv2013}, we show that for increasing number of particles $\Omega$
approaches $\sqrt{3}\omega_0$, the value predicted by Gross-Pitaevskii theory, for intermediate $U$.

Figure \ref{fig:nonmonotonicOmega} summarizes exact diagonalization results for $N = 2, \dots,7$,
for a fixed trapping strength $k=0.01$.  For larger particle numbers, this means significant
deviation from the low-density limit where the continuum limit (effective mass approximation) is
valid, as seen from the values of the central density in Figure \ref{fig:nonmonotonicOmega}b.  At
$U\to0$, the breathing frequency deviates slightly from the continuum value $\Omega=2\omega_0$ but
is the same for all $N$, as explained in Sec.\ \ref{sec_spectrum_free_points}, despite the increase
of the central density with $N$.  In the $U\to\infty$ limit, the deviation from the continuum
$\Omega=2\omega_0$ is stronger for larger $N$.  Since we work with finite chains, there is also some
deviation visible at large $U$ from the single-particle predictions with infinite lattices.  (The
predictions are listed on the right of Figure \ref{fig:nonmonotonicOmega}a.)  This is because the
additional confinement due to a finite lattice affects the spectrum, as noted previously in Figure
\ref{fig:spectrum1}.  The effect is noticeable even though the ground-state occupancy of the edge
site is less than $10^{-3}$ (Figure \ref{fig:nonmonotonicOmega}d). For example, at $U = 102$ the
breathing frequency is $\left(\Omega/\omega_0\right)^2 = 3.50 $ for $N=7$ and $L=36$, which is
larger than $3.46$, the value calculated from the single-particle spectrum in an infinite lattice
with a $k=0.01$ trap.

Despite these ``lattice effects'' (finite-density effects) at small and large $U$, the non-monotonic
behavior of the $\Omega$ versus $U$ curve (Figure \ref{fig:nonmonotonicOmega}a) reflects the overall
continuum expectation outlined in the Introduction and explored in
Ref.\ \cite{KroenkeSchmelcher_arxiv2013}.  For each $N$, there is a prominent minimum of $\Omega$ at
a finite $U$.  The minimum value is plotted in Figure \ref{fig:nonmonotonicOmega}e; the available
data strongly indicates that the minimum value of $\Omega$ goes to $\sqrt{3}\omega_0$ as $N$ is
increased at fixed $k$.  The position of the minimum moves to smaller $U$ with increasing number
(Figure \ref{fig:nonmonotonicOmega}d).

An obvious question is whether at larger $N$ the structure remains a pronounced minimum or whether
it becomes a broad plateau at or near $(\Omega/\omega_0)^2=3$.  Figure \ref{fig:nonmonotonicOmega}f
provides a partial answer.  We note that the large-$N$ limit may be taken in inequivalent ways on
the lattice.  In Figure \ref{fig:nonmonotonicOmega}a-e, we have kept $k$ constant.  It is also
reasonable to vary $N$ with constant central density, which requires adjusting $k$ for each $N$ and
$U$.  Figure \ref{fig:nonmonotonicOmega}f shows results for both these schemes.  We plot the rate of
decrease $c_1$ of $\Omega$ with $U$, at small $U$.  This is extracted by fitting the $\Omega(U)$
function for $U\lesssim0.15$ with a third-order polynomial; $c_1$ is the linear coefficient of this
fit.  In both cases (constant $k$ and constant $n_{L/2}$), the slope magnitude increases linearly
with the number of bosons $N$.  This indicates that, with large enough $N$, the breathing frequency
goes down from the non-interacting value $(\Omega/\omega_0)^2=4$ toward the mean-field value
$(\Omega/\omega_0)^2=3$ at very small interactions, so that at larger $N$ the $\Omega(U)$ curve
should show a broad valley rather than a sharp minimum.  This is consistent with the continuum
results of Ref.\ \cite{KroenkeSchmelcher_arxiv2013}.

\section{Failure of Gutzwiller approximation \label{sec:Gutzwiller}}

In this section we use the Gutzwiller mean field approximation for the Bose-Hubbard model
\cite{Gutzwiller_lewensteinreview, Gutzwiller_various} to calculate the time evolution after a trap
quench and hence obtain the breathing mode frequency.  We show that the Gutzwiller approximation
fails dramatically to reproduce the breathing mode frequency at large $U$.

The Gutzwiller approximation uses the product wave function $ |\Psi(t)\rangle = \prod_i \sum_n
f_n^{(i)}(t) |i,n\rangle, $ where $i$ is the site index and $n$ indicates a single site number state.
The time-dependent coefficients $f_n^{(i)}(t)$ describe the evolution of the
system.  This ansatz captures local number fluctuations but ignores correlations between sites.  
The dynamics is governed by the coupled differential equations
\begin{equation}
\begin{split}
  i\frac{\text{d} f_n^{(i)}}{\text{d} t} = & -\sum_{\langle i,j \rangle} \left( \sqrt{n+1} \Phi_j^* f_{n+1}^{(i)} + \sqrt{n} \Phi_j f_{n-1}^{(i)} \right) \\
 	      & + \left( \frac{U}{2} n(n-1) + V(i)  \right) f_{n}^{(i)}, 
\end{split}
\end{equation}
where $\Phi_i = \langle b_i \rangle = \sum_n \sqrt{n} (f_{n-1}^{(i)})^*f_{n}^{(i)}$.

\begin{figure}[t]
\centering
  	\includegraphics[width=1\columnwidth]{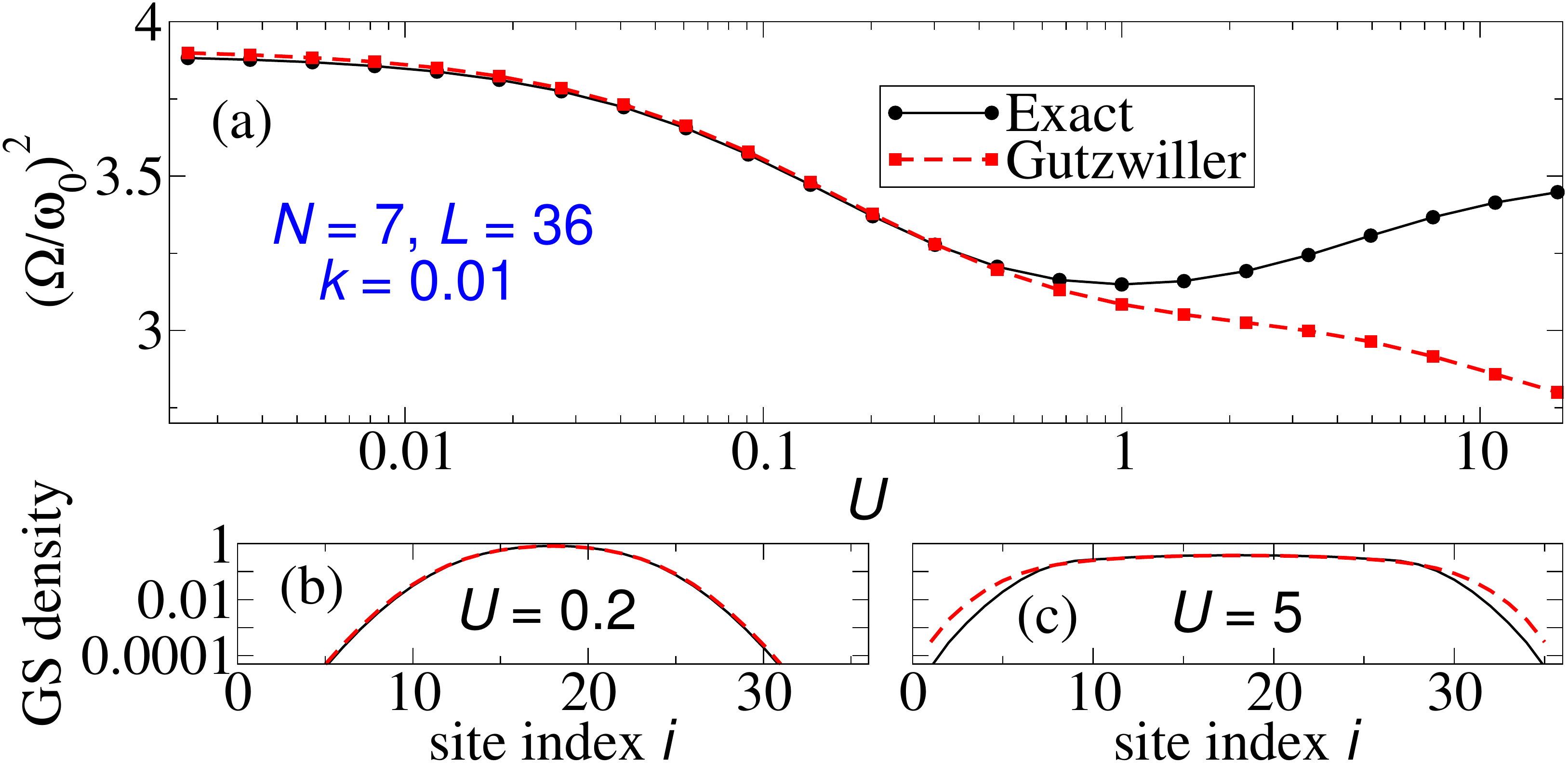}
\caption{(color online) 
(a) Comparison between the exact breathing frequency (obtained from exact diagonalization of the
  Hamiltonian) and the Gutzwiller approximation (obtained from $I(t)$ oscillations in real-time
  Gutzwiller evolution).  For $U\gtrsim1$, the Gutzwiller result shows a decrease of $\Omega$, in
  contradiction to an exact result.
(b,c) Ground state density for $k=0.01$, for (b) $U=0.2$ and (c) $U=5$. At large $U$
  the ground state density given by the Gutzwiller approximation deviates from the exact
  result.
}
\label{fig:EDGutz}
\end{figure}

Figure \ref{fig:EDGutz}a shows a comparison of the breathing frequency obtained from a trap quench
performed using the Gutzwiller approximation, with the exact value calculated from exact
diagonalization of the Hamiltonian.  For small $U$ the Gutzwiller approximation reproduces well the
decrease of $\Omega/\omega_0$ from $2$ toward $\sqrt{3}$.  However, for $U > 1$ Gutzwiller shows
further decrease of the breathing frequency with increasing $U$.    

In Figures \ref{fig:EDGutz}(b,c) we compare ground states obtained with Gutzwiller theory compared to
the exact ground states.  There are deviations at large $U$, highlighted through the log-linear
scale.  Since the same breathing mode frequencies are obtained in different quench strengths, we
surmise that the qualitative failure at large $U$ is not due to the initial, but rather due to some
fundamental shortcoming of the Gutzwiller approximation which, to the best of our knowledge, is not
yet well understood.  

One way of viewing this discrepancy is that the Gutzwiller approximation fails to describe the
fermionization of the system.
However, it is likely that this failure is not a 1D feature only,  as the Gutzwiller approximation
has been found to incorrectly predict the expansion speed after release from a trap, in both 1D and
higher dimensions \cite{Schneider_Heidrich-Meisner_expansion_PRL2013}.

\section{Lattice effects at strong confinement  \label{sec:strongtrap}}

\begin{figure}
\centering
  	\includegraphics[width=1\columnwidth]{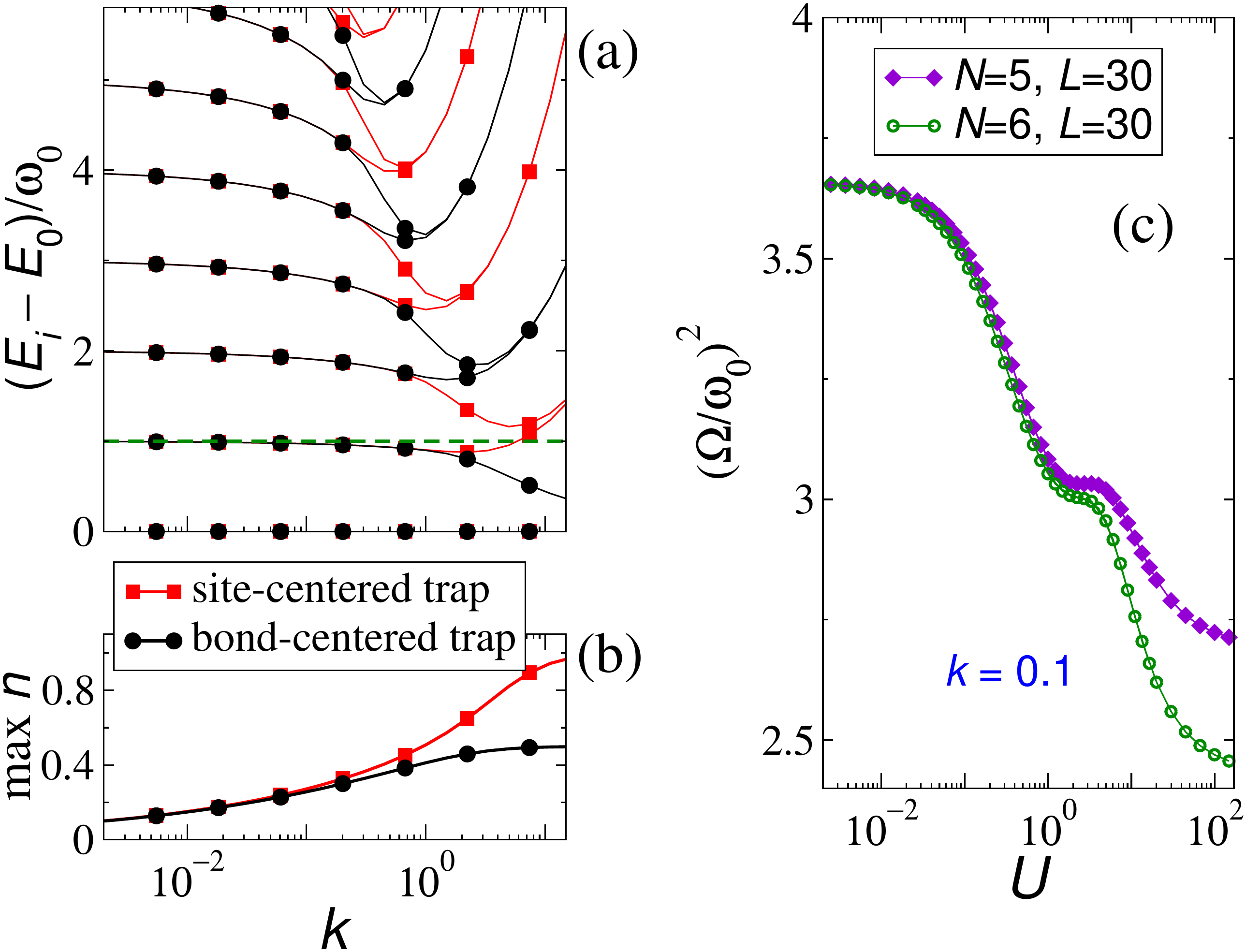}
\caption{\label{fig:appendix1} (color online) (a) Single particle energy spectrum as a function of
  trapping strength for site-centered trap (odd-sized chain) and bond-centered trap (even-sized
  chain).  At larger $k$, the first excited level (dipole mode) deviates from $\omega_0=\sqrt{2k}$,
  shown as dashed (green) horizontal line.  (b) Maximum (central) density of single-particle ground
  state.  (c) Breathing frequency at relatively strong trapping potential $k=0.1$. For $U\gtrsim1$,
  $\left(\Omega/\omega_0\right)^2$ drops below 3.  }
\end{figure}

The focus of this work has been on the low-density limit or weak trapping, where comparison to
continuum physics is meaningful.  In this section we describe some effects of strong trapping
situations where the bosons are confined to a few lattice cites.

Figure \ref{fig:appendix1}a shows the single particle spectrum as a function of trapping strength
$k$. The energy spectrum gets denser with increasing $k$ as one moves away from the continuum-like
limit where the levels are equally spaced.  The energy gaps between successive eigenstates decrease
with increasing $k$, and this effect is stronger for the higher energy levels.  For large $k$ the
ground state wavefunction is spread only over a few lattice sites with a high central density
(Figure \ref{fig:appendix1}b).  The deviation from continuum physics, or sensitivity to lattice
details, also shows up in the difference between a trap centered at a site (site-centered, odd
lattice) and a trap centered between two sites (bond-centered, even lattice).  (In our scheme of
placing the trap center at $(L+1)/2$, this is the difference between odd and even sized lattices.)
For $k\to\infty$ the spectrum is degenerate, except for the ground state in the site-centered
case.   The failure of the effective mass approximation is also seen through the first excitation
energies deviating from  $\omega_0=\sqrt{2k}$ as $k$ is increased (Figure \ref{fig:appendix1}a).

For larger $k$, as a function of the interaction strength $U$, the breathing frequency (Figure
\ref{fig:appendix1}c) decreases to $\Omega\sim\sqrt{3}\omega_0$ as $U$ is increased up to $U\sim1$.
At stronger interactions, the breathing frequency \emph{decreases} further, unlike the continuum or
small $k$ situations.  This can be understood in the large $U$ (free fermion) limit: since the
single-particle spectrum is dense for $k\sim0.1$ (Figure \ref{fig:appendix1}a), the difference
between the $N$-th and $(N+2)$-th levels levels is smaller than $\sqrt{3}\omega_0$.  There is an
interesting plateau at $\Omega\sim\sqrt{3}\omega_0$ ($U\sim1$) which is difficult to explain from
single-particle considerations.

\section{Summary and Discussion}

We have presented a detailed study of breathing modes for bosons on a tight-binding finite chain
subject to a harmonic trap, and repulsively interacting through on-site interactions, as described
by the 1D Bose-Hubbard model.  We have used exact numerical diagonalization to treat up to 7 bosons,
and used the excitation spectrum, overlaps after a quench, and explicit time evolution, to clarify
various aspects of the problem.  The continuum calculations of
Ref.\ \cite{KroenkeSchmelcher_arxiv2013} are similar in spirit.  In addition to connecting to the
continuum limit, we have also highlighted deviations from this limit.

As in the continuum, the breathing mode frequency turns out to be the excitation energy of  the
lowest excited reflection-symmetric state.  This is twice the trapping frequency, $2\omega_0$, at
small and large interactions, and it decreases to around $\sqrt{3}\omega_0$ at intermediate $U$.
There are various lattice corrections to this continuum picture, which we have examined in the
text, even for strong traps (Section \ref{sec:strongtrap}).   

Our results raise various open questions.  Most prominent is the failure of the time-dependent
Gutzwiller approximation to describe the breathing mode frequency at larger $U$ (Section
\ref{sec:Gutzwiller}).  Since this is a popular method for time evolution of Bose-Hubbard systems
\cite{Gutzwiller_various}, it is important to understand regimes and situations where it fails.  Our
results of Figure \ref{fig:EDGutz} should serve as a benchmark in further understanding of this
issue.   

It is possible to excite breathing modes through quenches of the on-site interaction $U$.  While we
expect essential features to be similar, it would be interesting to compare especially the overlap
profiles analogous to Figures \ref{fig:overlaps_N567} for the trap quench.

We have restricted ourselves to low-density situations where Mott physics does not play a role.
Clearly, dynamics in the presence of one or multiple Mott cores is an intriguing direction of study,
especially if it is possible to go beyond Gutzwiller dynamics.


\begin{thebibliography}{99}

\bibitem{early_BEC} 
D.~S.~Jin, J.~R.~Ensher, M.~R.~Matthews, C.~E.~Wieman, and E.~A.~Cornell, Phys.\ Rev.\ Lett.
 \textbf{77}, 420 (1996).
%
D.~M.~Stamper-Kurn, H.-J.~Miesner, S.~Inouye, M.~R.~Andrews, and W.~Ketterle,
Phys.\ Rev.\ Lett.\ \textbf{81}, 500 (1998). 
%
F.~Dalfovo, S.~Giorgini, L.~P.~Pitaevskii, and S.~Stringari, Rev.\ Mod.\ Phys.\ \textbf{71}, 463 (1999). 
%
C.~J.~Pethick and H.~Smith, \textit{Bose-Einstein Condensation in Dilute Gases}, Cambridge
University Press; 1st edition (2002); 2nd edition (2008).


\bibitem{Naegel_exp} E.~Haller, M.~Gustavsson, M.~J.~Mark, J.~G.~Danzl, R.~Hart, G.~Pupillo,
  H.~C.~N\"agerl, Science \textbf{325}, 1224 (2009).  


\bibitem{Snoek_PRA12} M.~Snoek, Phys. Rev. A \textbf{85}, 013635 (2012). 

\bibitem{exactSolutionTrap} A.~M.~Rey, G.~Pupillo, C.~W.~Clark, and C.~J.~Williams, Phys.\ Rev.\ A
  \textbf{72}, 033616 (2005).  

\bibitem{Lundh} E.~Lundh, Phys.\ Rev.\ A \textbf{70}, 033610 (2004); Phys. Rev. A \textbf{70},
  061602(R) (2004). 

\bibitem{Peotta_DiVentra_2013} S.~Peotta and M.~Di~Ventra,  arXiv:1303.6916. 
      
\bibitem{Caux_Konnik_several} 
G.~Brandino, J.-S.~Caux, and R.~M.~Konik,  arXiv:1301.0308.
%
J.-S.~Caux and R.~M.~Konik, Phys.\ Rev.\ Lett.\ \textbf{109}, 175301 (2012).   


\bibitem{ZollnerSchmelcher_PRA07}
S.~Z\"ollner, H.-D.~Meyer, and P.~Schmelcher, Phys.\ Rev.\ A \textbf{75}, 043608 (2007). 


\bibitem{Brouzos} I.~Brouzos and P.~Schmelcher, Phys. Rev. A \textbf{85}, 033635 (2012).

\bibitem{KroenkeSchmelcher_arxiv2013}
R.~Schmitz, S.~Kr\"onke, L.~Cao, and P.~Schmelcher, arXiv:1306.5665. 

\bibitem{CederbaumAlon_arxiv2013}
J.~Grond, A.~I.~Streltsov, A.~U.~J.~Lode, K.~Sakmann, L.~S.~Cederbaum, and O.~E.~Alon, 
arXiv:1307.1667.

\bibitem{Astrakharchik} G.~E.~Astrakharchik, J.~Boronat, J.~Casulleras, and S.~Giorgini,
  Phys.\ Rev.\ Lett.\ \textbf{95}, 190407 (2005). 
  
\bibitem{Menotti} 
C.~Menotti and S.~Stringari, Phys.\ Rev.\ A \textbf{66}, 043610 (2002).

\bibitem{Kraemer_Pitaevskii_Stringari_PRL02}
M.~Kraemer, L.~Pitaevskii, and S.~Stringari, Phys.\ Rev.\ Lett.\ \textbf{88}, 180404 (2002). 

\bibitem{Kimura_PRA02} T.~Kimura,  Phys.\ Rev.\ A \textbf{66}, 013608 (2002).

\bibitem{FuchsLeyronasCombescot_2003_2004} J.~N.~Fuchs, X.~Leyronas, R.~Combescot, 
Phys.\ Rev.\ A \textbf{68}, 043610 (2003); Laser Physics \textbf{14},1 (2004).

\bibitem{Mazets_EPJD11}  
I.~E.~Mazets, Eur.\ Phys.~J.~D \textbf{65}, 43 (2011).

 
\bibitem{Esslinger_exp} H.~Moritz, T.~St\"oferle, M.~K\"ohl, and T.~Esslinger,
  Phys.\ Rev.\ Lett.\ \textbf{91}, 250402 (2003).   

\bibitem{Stringari} S.~Stringari, Phys.\ Rev.\ A \textbf{58}, 2385 (1998).
  
  
\bibitem{HaugsetHaugerud_PRA98} T.~Haugset and H.~Haugerud,  Phys.\ Rev.\ A \textbf{57}, 3809 (1998).  


\bibitem{integer_partitions} The function $p(n)$ generates the well-known sequence
  1,1,2,3,5,7,11,15,... counting the number of integer partitions of the integer $n$.  This is listed
  as sequence A000041 in the Online Encyclopedia of Integer Sequences,
  \href{http://oeis.org}{http://oeis.org}.

\bibitem{restricted_integer_partitions} The number of partitions of the integer $n$ into a maximum
  of $N$ parts, $p_N(n)$, differs from the unrestricted sequence $p(n)$ at larger $n$.  For $N=$
  2,3,4,5,6,7,8, the sequences $p_N(n)$ are listed as A008619, A001399, A001400, A001401, A001402,
  A008636, A008637, in the Online Encyclopedia of Integer Sequences,
  \href{http://oeis.org}{http://oeis.org}.


\bibitem{BongsSengstockPfannkuche_PRA07}  F.~Deuretzbacher, K.~Bongs, K.~Sengstock, and
D.~Pfannkuche, Phys.\ Rev.\ A \textbf{75}, 013614 (2007). 
    
\bibitem{KohnMode} L.~Brey, N.~F.~Johnson, and B.~I.~Halperin, Phys.\ Rev.\ B \textbf{40}, 10647
  (1989).  M.~Bonitz, K.~Balzer, and R.~van~Leeuwen, Phys. Rev. B \textbf{76}, 045341 (2007). 
  
  
\bibitem{Gutzwiller_lewensteinreview} 
%
The Gutzwiller approximation is reviewed in 
Section 3.4 of: 
M.~Lewenstein, A.~Sanpera, V.~Ahufinger, B.~Damski, A.~Sen (De), and U.~Sen,
Adv.\ Phys.\ {\bf 56}, 243 (2007).  

\bibitem{Gutzwiller_various} 
%
D.~S.~Rokhsar and B.~G.~Kotliar, Phys.\ Rev.\ B {\bf 44}, 10328 (1991).
%
W.~Krauth, M.~Caffarel, and J.-P.~Bouchaud, Phys.\ Rev.\ B {\bf 45}, 3137 (1992).
%
K.~Sheshadri, H.~R.~Krishnamurthy, R.~Pandit, and T.~V.~Ramakrishnan,
Europhys.\ Lett.\ {\bf 22}, 257 (1993).
%
D.~Jaksch, C.~Bruder, J.~I.~Cirac, C.~W.~Gardiner, and P.~Zoller,
Phys.\ Rev.\ Lett.\ {\bf 81}, 3108 (1998).
%
D.~Jaksch, V.~Venturi, J.~I.~Cirac, C.~J.~Williams, and P.~Zoller, Phys.\
Rev.\ Lett.\ {\bf 89}, 040402 (2002).
%
B.~Damski, J.~Zakrzewski, L.~Santos, P.~Zoller, and M.~Lewenstein, Phys.\
Rev.\ Lett.\ {\bf 91}, 080403 (2003). 
%
J.~Zakrzewski, Phys.\ Rev.\ A {\bf 71}, 043601 (2005).
%
X.~Lu and Y.~Yu, Phys.\ Rev.\ A {\bf 74}, 063615 (2006).
%
M.~Snoek and W.~Hofstetter, Phys.\ Rev.\ A {\bf 76}, 051603(R) (2007).
%
C.~Trefzger, C.~Menotti, and M.~Lewenstein, Phys. Rev. A 78, 043604 (2008). 
%
P.~Navez and R.~Schützhold, Phys.\ Rev.\ A {\bf 82}, 063603 (2010).
%
S.~Natu, K.~Hazzard, and E.~Mueller, Phys.\ Rev.\ Lett.\ {\bf 106}, 125301 (2011).

  

\bibitem{Schneider_Heidrich-Meisner_expansion_PRL2013}
J.~P.~Ronzheimer, M.~Schreiber, S.~Braun, S.~S.~Hodgman, S.~Langer, I.~P.~McCulloch,
F.~Heidrich-Meisner, I.~Bloch, and U.~Schneider,  Phys.\ Rev.\ Lett.\ \textbf{110}, 205301 (2013).     


\end{thebibliography}
\end{document}